\documentclass[pre,twocolumn,showpacs,superscriptaddress,floatfix]{revtex4}
\usepackage[latin1]{inputenc}
\usepackage{graphicx}
\usepackage{lipsum}
\usepackage{amsmath}
\usepackage{amssymb}
\usepackage{pifont}
\usepackage{epstopdf}
 \usepackage{color}

\begin{document}
\title{Flows and mixing in channels with misaligned superhydrophobic walls
}

 \author{Tatiana V. Nizkaya}
\thanks{These three authors contributed equally to the work.}
  \affiliation{A.N. Frumkin Institute of Physical Chemistry and
Electrochemistry, Russian Academy of Science, 31 Leninsky Prospect,
119991 Moscow, Russia}

\author{Evgeny S. Asmolov}
\thanks{These three authors contributed equally to the work.}
\affiliation{A.N. Frumkin Institute of Physical Chemistry and
Electrochemistry, Russian Academy of Science, 31 Leninsky Prospect,
119991 Moscow, Russia}
\affiliation{Central Aero-Hydrodynamic Institute, 140180 Zhukovsky,
Moscow region, Russia}

\author{Jiajia Zhou}
\thanks{These three authors contributed equally to the work.}
\affiliation{Institut f\"ur Physik, Johannes Gutenberg-Universit\"at
 Mainz, D55099 Mainz, Germany}

 \author{Friederike Schmid}
 \affiliation{Institut f\"ur Physik, Johannes Gutenberg-Universit\"at
   Mainz, D55099 Mainz, Germany}
 \author{Olga I. Vinogradova}
\email[Corresponding author: ]{oivinograd@yahoo.com}
 \affiliation{A.N. Frumkin Institute of Physical Chemistry and
   Electrochemistry, Russian Academy of Science, 31 Leninsky Prospect,
   119991 Moscow, Russia}
 \affiliation{Department of Physics, M.V. Lomonosov Moscow State
   University, 119991 Moscow, Russia}
 \affiliation{DWI - Leibniz Institute for Interactive Materials, RWTH Aachen, Forckenbeckstr. 50, 52056 Aachen,
   Germany}

\date{\today}

\begin{abstract}
Aligned superhydrophobic surfaces with the same texture orientation reduce drag
in the channel and generate secondary flows transverse to the direction of the
applied pressure gradient.  Here we show that a transverse shear
can be easily generated by using superhydrophobic channels with misaligned textured surfaces.
 We propose a general theoretical approach to quantify this transverse flow by
introducing the concept of an effective shear tensor.  To illustrate its use,
we present approximate theoretical solutions and Dissipative Particle Dynamics
simulations for striped superhydrophobic channels. Our results demonstrate that
the transverse shear leads to complex flow patterns, which provide a new mechanism of a passive vertical mixing at the scale of a
texture period. Depending on the value of Reynolds number two different scenarios occur. At relatively low Reynolds number
the flow represents a transverse shear superimposed with two co-rotating vortices.
 For larger Reynolds number these vortices become isolated, by suppressing fluid transport in the transverse direction.
\end{abstract}


\maketitle

\section{Introduction}
\label{sec:introduction}

Superhydrophobic (SH) textures have raised a considerable interest and
motivated numerous studies during the past decade~\cite{quere.d:2005,darmanin.t:2014}. Such surfaces in the Cassie
state, i.e., where the texture is filled with gas, are extremely important in
the context of fluid dynamics due to their superlubricating
potential~\cite{bocquet2007,rothstein.jp:2010,vinogradova.oi:2012}. The use of
highly anisotropic SH textures provides additional possibilities for flow
manipulation. The effective hydrodynamic slip of such surfaces, $\mathbf
b_{\rm eff}$, is generally
tensorial~\cite{stone2004,Bazant2008,feuillebois.f:2009,harting.j:2012}  due to secondary
flows transverse to the direction of the applied pressure
gradient~\cite{mixer2010,vinogradova.oi:2010}. This can be used to separate particles~\cite{pimponi.d:2014} and enhance mixing rate~\cite{ou.j:2007} in typical for microfluidic devices
low-Reynolds-numbers flows. Over the last decade, the quantitative understanding of liquid flow in SH
channels with anisotropic walls was significantly expanded. However, many fundamental issues still remains challenging.

According to the modern concept of effective slip, $\mathbf b_{\rm eff}$
is a global characteristic of a channel~\cite{vinogradova.oi:2012}, which can
be applied for arbitrary channel thickness~\cite{harting.j:2012}.
This implies that the eigenvalues depend
not only on the parameters of the heterogeneous surfaces (such as local slip
lengths, fractions of phases, and the texture period $L$), but also on the
channel thickness $H$. However, for a thick (compared to $L$) channel they
become a characteristics of a heterogeneous interface
solely~\cite{Bazant2008,bocquet2007,kamrin.k:2010}.

\begin{figure}[t]
\vspace*{-0.1\columnwidth} \hspace*{-0.1\columnwidth}
\includegraphics[width=\columnwidth]{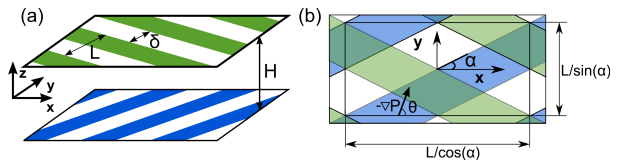}
\caption{(a) Cartoon showing the two identical misaligned superhydrophobic
walls decorated with no-slip (white) and partial-slip (colored) stripes. (b)
Top view of a periodic rectangular unit cell.}
\label{fig:sketch}
\end{figure}

Most of the prior work has focused on the optimization of the (forward)
effective slip and calculations of the eigenvalues of the effective
slip-length tensor for a single 1D interface
~\cite{Belyaev2010,Zhou2013,priezjev.n:2011,Ybert2007,ng.co:2010} and channels
with aligned walls~\cite{feuillebois.f:2009,Zhou2012} or one hydrophilic
(no-slip) wall~\cite{feuillebois.f:2009,harting.j:2012}. The eigenvalues of
$\mathbf b_{\rm eff}$ correspond to the directions of fastest and lowest
effective slip. In these directions, a pressure gradient does not produce
transverse flows. In other directions, however,  a flow may become
misaligned with the force, and transverse hydrodynamic flow can be generated.
This phenomenon was also discussed only for a single
interface~\cite{Bazant2008,vinogradova.oi:2012}, for channels with
symmetrically aligned stripes~\cite{mixer2010,Zhou2012} or with one
no-slip wall~\cite{mixer2010,nizkaya.tv:2013}.

In the present paper, we show that any
misalignment of textured and anisotropic walls necessarily leads to a generation of shear flow in the transverse
direction, which has not
been appreciated in previous work~\cite{Bazant2008}.  The transverse shear, in turn, generates complex flow structures at the
scale of the texture period.We show that at low, but finite, Reynolds
numbers (Re) of the channel, a transverse shear is superimposed with two co-rotating vortices. At larger Re, these vortices become isolated, by suppressing a transverse shear. Such flow structures lead to a global vertical mixing of
fluids, which does not rely on the presence of confining side walls.
\section{Model}
\label{sec:model}

We consider the pressure-driven
flow between two parallel
stationary (passive) SH walls separated by the distance $H$ and decorated with
identical anisotropic textures of a period $L$, and width of the gas area $\delta$, so that the fraction of the slipping area is $\phi=\delta/L$. The lower and the upper wall textures are misaligned by an angle $0\leq 2\alpha \leq \pi/2$. The flow is periodic with a rectangular unit cell as depicted in
Fig. \ref{fig:sketch}.  We place the origin of the $(x,y,z)$ coordinate
system at the center of the cell, in the midplane, with the $z$ axis
perpendicular to the walls.

We assume the gas interface to be flat with no
meniscus curvature. Such an idealized  situation, which neglects an additional mechanism for a dissipation due to a meniscus, has been considered in most
previous publications~\cite{priezjev.nv:2005,Ybert2007,vinogradova.oi:2010}. It has been shown theoretically~\cite{Ybert2007,Davis:2009} that a low curvature of the meniscus expected for relatively dense textures  cannot reduce the effective slippage in the Stokes regime (but note that there are still some remaining controversies in the experimental data~\cite{ou2004,Bolognesi:2014}). The quantitative understanding of whether and how such curvature effects could modify effective slip lengths in the case of finite Re is still challenging. This subject, which remains largely unexplored, is beyond the scope of the present work and deserves a separate study.

We finally impose no-slip at the solid area, i.e. neglect slippage of liquid~\cite{vinogradova1999,vinogradova.oi:2009,joly.l:2006} and gas~\cite{seo.d:2013} past hydrophobic surface, which is justified provided the nanometric slip is
small compared to parameters of the texture. Prior work often assumed idealized shear-free boundary
conditions over the gas sectors~\cite{philip.jr:1972,priezjev.nv:2005,lauga.e:2003}, so that the viscous dissipation in the underlying gas phase has been neglected. Here we
will use the partial slip boundary conditions, which are the consequences of the `gas cushion model'~\cite{vinogradova.oi:1995a,nizkaya.tv:2014}.


\section{Theory of transverse shear}
\label{sec:theory}

\begin{figure}
  \includegraphics[width=0.5\columnwidth]{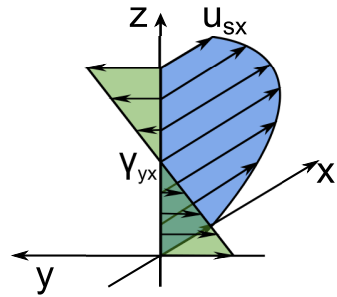}
  \caption{Sketch of an mean velocity profile $\left\langle u \right\rangle (z)$ given by Eq. (\protect\ref{u_me}). A uniform $- \nabla P$ is applied in the $x-$direction. }
  \label{fig:sketch2}
\end{figure}

To address effective (`macroscopic') properties of the channel we evaluate the mean
velocity profile of the Stokes flow (Re $\ll 1$), averaged over the periodic cell in $x,y$. Due to the
linearity of the Stokes equations, it is sufficient to consider only the flow
in eigendirections of the texture. Let us first consider a pressure-driven flow with $-\nabla P$ aligned with the $x-$axis ($\theta=0$).
If $-\nabla P$ is aligned with the $y-$axis, the flow can then be obtained by simply replacing $\alpha$ by $\pi/2-\alpha$. The symmetry of the problem implies that the $x$-component of
the mean velocity is symmetric in $z$ and  its $y$-component is
antisymmetric. 
For the Stokes flow the mean velocity profile is scaled by
$U_0=-H^{2}\mathbf{\nabla }P/\left( 2\mu \right)$ and has the following form:

\begin{equation}
\left\langle \mathbf{u}\right\rangle (z)=\left( \frac{1}{4}-z^{2}\right) \mathbf{e}_{x}+u_{sx}%
\mathbf{e}_{x}+\gamma _{yx}z\mathbf{e}_{y},
\label{u_me}
\end{equation}%
where coordinates are scaled by $H$, $u_{sx}$ is the slip velocity in the $x-$direction and $\gamma_{yx}$ is the transverse shear rate. Therefore, the first term  represents the conventional Poiseuille flow, the second term is a slip-driven (forward) plug flow, and the last term is a linear shear flow in the transverse direction. The laminar flow in the channel
is a linear superposition of these
three terms as shown in Fig.~\ref{fig:sketch2}.  Eq.(\ref{u_me}) can
therefore be seen as a generalization of an earlier idea, formulated for a
thick channel~\cite{Bazant2008}, to an arbitrary channel situation.

To quantify the effective properties of the channel, we
now introduce the flow and shear rates averaged over the cell volume $V$:%
\begin{eqnarray}
\mathbf{Q} &=&\frac{1}{V}\int \mathbf{u}_{\tau }dV=-\frac{H^{2}}{12\mu }%
\mathbf{k}\cdot \mathbf{\nabla }P, \\
\mathbf{G} &=&\frac{1}{V}\int \frac{\partial \mathbf{u}_{\tau }}{\partial z}%
dV=-\frac{H}{2\mu }\boldsymbol\gamma\cdot \mathbf{\nabla }P,
\label{qg}
\end{eqnarray}
where $\mathbf{u}_{\tau }=\left( u_{x},u_{y}\right) $ is the tangential
velocity. The effective (dimensionless) effective permeability, $\mathbf{k},$ and shear,
$\boldsymbol\gamma,$  tensors can then be found by solving a local problem in
the channel.

Let us note that Eq.~(\ref{u_me}) and the symmetry conditions imply that
the $\mathbf{k}$ and $\boldsymbol\gamma$ tensors   are:
\begin{equation}
\mathbf{k}=\left(
\begin{array}{cc}
k_{xx}(\alpha) & 0 \\
0 & k_{yy}(\alpha)%
\end{array}%
\right) , \;
\end{equation}
where $k_{xx}=1+6u_{sx}$ and
\begin{equation}
\quad \boldsymbol\gamma=\left(
\begin{array}{cc}
0 & \gamma _{xy}(\alpha) \\
\gamma _{yx}(\alpha) & 0%
\end{array}%
\right).
\label{kgam}
\end{equation}
Thus, our coordinate system diagonalizes the permeability tensor.
Also, due to the symmetry of the periodic cell,
$k_{yy}(\alpha)=k_{xx}(\pi/2-\alpha)$ and
$\gamma_{xy}(\alpha)=\gamma_{yx}(\pi/2-\alpha)$.

To calculate $u_{sx}(\alpha)$ and $\gamma_{yx}(\alpha)$ we have to solve the
Stokes equations on the periodic cell with spatially varying local slip lengths. The flow can be treated as a superposition of
two solutions for a configuration with one SH and one hydrophilic
wall, which is justified provided the thickness of the
channel is large enough (see Appendix~\ref{appA}). This simplification allows us to obtain explicit formulae for the
mean slip and the shear tensor:
\begin{eqnarray}
u_{sx}(\alpha)=\frac{\beta_++\beta_-\cos(2\alpha)+2(\beta_+^2-\beta_-^2)}{1+2\beta_+-2\beta_-\cos(2\alpha)},\\
\gamma_{yx}(\alpha)=-\frac{2\beta_-\sin(2\alpha)}{1+2\beta_+-2\beta_-\cos(2\alpha)},
\end{eqnarray}%
where $\beta_{+}=(b_{\mathrm{eff}}^{\Vert }+b_{\mathrm{eff}}^{\perp })/(2H)$,
$\beta_{-}=(b_{\mathrm{eff}}^{\Vert }-b_{\mathrm{eff}}^{\perp })/(2H).$ The
effective slip tensor $\mathbf b_{\mathrm{eff}}$ here corresponds to a
configuration  with one SH wall.

The misalignment of the upper and the lower textures leads to a `macroscopic'
anisotropy of the flow, i.e. the effective permeability depends on the direction of
the pressure gradient, and a secondary transverse flow is generated.
To study this effect, we now apply $-\nabla P$
at an angle $0 \leq \theta \leq \pi/2$ to the $x$-axis.
For the forward and transverse flow rate we then obtain, using the tensorial formalism~\cite{Bazant2008}:
\begin{equation}
\begin{array}{ll}
\displaystyle Q_{\mathrm{F}}/Q_0=k_{xx}\cos^2\theta+k_{yy}\sin^2\theta,
\label{QL} \\
\displaystyle Q_{\mathrm{T}}/Q_0=(k_{xx}-k_{yy})\cos\theta\sin\theta,
\end{array}
\end{equation}
where $Q_0=\dfrac{|\nabla P|H^2}{12\mu}$ corresponds to the Poiseuille flow in a channel with hydrophilic walls. Note that $Q_{\mathrm{T}}$ vanishes when $\theta = 0$ or when stripes are crossed at a right angle, $2 \alpha = \pi/2$. These special configurations, therefore, are equivalent to channels with impermeable side walls.

Similarly, for the mean shear rates we get

\begin{equation}
\begin{array}{ll}\displaystyle G_{\mathrm{F}}/G_0=-(\gamma_{xy}+\gamma_{yx})\sin\theta\cos\theta,
\label{GL} \\
\displaystyle G_{\mathrm{T}}/G_0=\gamma_{yx}\cos^2\theta-\gamma_{xy}\sin^2\theta,
\end{array}
\end{equation}
where $G_0=\dfrac{U_0}{H}=\dfrac{|\nabla P| H}{2\mu}$ is a characteristic shear
rate for a hydrophilic channel. It follows from Eq.~(\ref{GL})  that the forward shear rate, $G_{\mathrm{F}}$, vanishes when  $\theta = 0$ or when $\gamma_{xy}=-\gamma_{yx}$, but
the transverse shear rate, $G_{\mathrm{T}}$, becomes zero only at $\theta
=\pm (\arctan(\gamma_{xy} /\gamma_{yx} ))^{1/2}$. However, this does not imply
that the shear tensor can be diagonalized by a proper rotation of the coordinate
system, since the two directions are not orthogonal in the general case.

To illustrate this general approach we focus below on specific case of SH surfaces consisting of
periodic stripes, where the piecewise constant local slip lengths of (identical) textured walls vary in only one direction. In this case the $\mathbf{k}$ and $\boldsymbol
\gamma$ tensors  have been found using the numerical solutions of the Stokes
equations~\cite{nizkaya.tv:2013} (see Appendix \ref{appA} for details), and the eigenvalues of the slip-length tensor correspond to those calculated before~\cite{harting.j:2012}.

\section{Simulations and data analysis}
\label{sec:simulation}

In order to assess the validity of the above theoretical approach at the relatively low Re regime and to explore what happens
when Re is larger, we  employ the Dissipative Particle Dynamics (DPD) simulations \cite{Hoogerbrugge1992, Espanol1995} carried out using the open source package ESPResSo~\cite{ESPResSo}. Specifically,  we use DPD  as a Navier-Stokes solver based on a fluid of non-interacting particles \cite{Zhou2013} and combine that with a tunable-slip method~\cite{Smiatek2008} that allows one to vary the local slip length~\cite{Zhou2012, asmolov:2013, Zhou2013}.


Note that for misaligned stripes the simulation
box depends on $2\alpha$ due to the periodic
boundary conditions in the $x-$ and $y-$directions (see Fig.~\ref{fig:sketch}).  The three dimensions of the simulation
box are $(L/\sin\alpha, L/\cos\alpha, H+2\sigma)$, where $\sigma$ sets the length scale. The extra $2\sigma$ in
$z$-direction accounts for the depletion due to the excluded volume
interaction of the impermeable surfaces.
In the present study, we have used $L=20\,\sigma$, and
$H/L=1$ or $0.5$. All simulations have been performed for textures
with the fraction of gas phase $\phi=0.5$, where the Cassie state is typically stable. The local slip length on the slippery area was chosen to be $b/L=10$, which is close to the maximum attainable local slip length at the gas area, $b/L \simeq 16 \phi$~\cite{nizkaya.tv:2014}, in the case of deep textures, where again the Cassie state is  very stable.

 With a fluid density $\rho =
3.75\,\sigma^{-3}$, a typical system consists of $0.6 \text{--} 1.2\times 10^5$
particles.  The simulation starts with randomly distributed particles.  The
pressure-driven flow is modeled by applying  to all particles an external force in the $x-$direction.
The system is allowed to reach a steady state after $10^6$ time steps; then the
measurement is performed by averaging over $10^5$ time steps to obtain sufficient
statistics.
 We use a body force ranging from $f=0.004\varepsilon/\sigma$ to $0.04\varepsilon/\sigma$, where $\varepsilon$ sets the energy scale. These values of $f$ provide flows with Reynolds numbers varying from $\mathrm{Re}=2.8$ to $\mathrm{Re}=28$.
All simulations were performed with a time step $\Delta t = 0.01 \sqrt{m/\varepsilon}\sigma$, where $m$ is the mass of DPD particles, and the temperature of the system was set at $\varepsilon = k_BT$.
The DPD interaction parameter is chosen at $\gamma_{\rm DPD} = 5.0 \sqrt{m\varepsilon}/\sigma$ and the cutoff is $1.0\,\sigma$.
The shear viscosity is measured to be $\mu_s = 1.35 \pm 0.01 \sqrt{m\varepsilon}/\sigma^2$.

Mean fluxes and shear rates are calculated by averaging the simulation data over the periodic cell. The fluid velocity field obtained in simulations is used to compute a family of streamlines from the beginning to the end of the cell. These streamlines are then used to construct the vector field of fluid displacements over one texture period. Finally, to visualize the fluid  mixing we define a dense uniform grid in the channel cross-section at a distance of $N$ periods from the inlet, and for each point of this grid we then calculate backward streamlines to determine its initial position.



\section{Results and Discussion}
\label{sec:result}

\begin{figure}[h!]
  \includegraphics[width=1.\columnwidth]{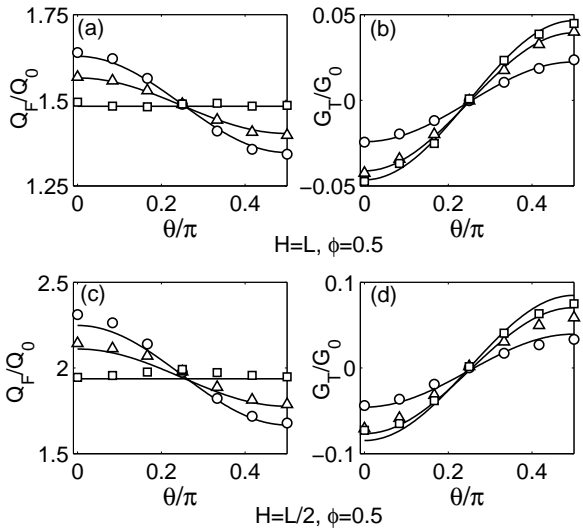}
   \caption{Forward flow rate (a, c) and transverse shear
rate (b, d) as a function of
$\theta$ computed with $b/L=10$. The thickness of the channel, $H,$ is equal to $L$ (a, b) or $L/2$ (c, d). Symbols show simulation results for Re$=2.8$ and
misalignment angles $2\alpha=\pi/6$ (circles), $\pi/3$ (triangles),
and $\pi/2$ (squares). Solid curves are theoretical predictions for Re $\ll 1$
(Eqs.~(\ref{QL}) and  (\ref{GL})). }
  \label{QGcurves}
\end{figure}	

In order to assess the validity of the above theoretical approach developed for low Re we first run a set of
simulations at Re=2.8 in channels with different misalignment angles $2\alpha$, and $H/L=1$ and $0.5$.
 The simulation results  are shown in Fig.~\ref{QGcurves}. Also included are the forward fluxes and transverse shear rates calculated using
Eqs.~(\ref{QL}-\ref{GL}).   A general conclusion is that the theoretical predictions are in excellent agreement with simulation results
for $H/L=1$
(Fig.~\ref{QGcurves}(a,b)). Even for thinner channels, $H/L=0.5$
(Fig.~\ref{QGcurves}(c,d)), where the theory slightly underestimates the permeability and
overestimates the transverse shear, which is likely due to simplifications made (see Section~\ref{sec:theory} and Appendix \ref{appA}), the agreement is quite good.
 Fig.~\ref{QGcurves} shows that the flow rates $Q_{\mathrm{F}}$
become maximal for  $-\nabla P$ aligned with the $x-$axis ($\theta = 0$) and minimal if it is aligned with the
$y$-axis ($\theta = \pi/2$). These directions  correspond to the largest amplitude of a transverse shear $|G_{\mathrm{T}}|$ at a given $2\alpha$. When the lower and upper stripes are orthogonal ($2\alpha=\pi/2$),  $Q_{\mathrm{F}}$ becomes independent on $\theta$, and we recall that in this case $Q_{\mathrm{T}}=0$ as discussed above. This orthogonal configuration allows one to  reach (at $\theta = 0$ and $\pi/2$) the largest possible amplitude of $|G_{\mathrm{T}}|$ that can be attained in the channel. Therefore, the detailed investigation of flow and mixing properties below will be restricted to this optimal configuration with  $2 \alpha=\pi/2$, $\theta=0$, and to a thinner channel, $H/L=0.5$.

\begin{figure}[t]
\includegraphics[width=1\columnwidth]{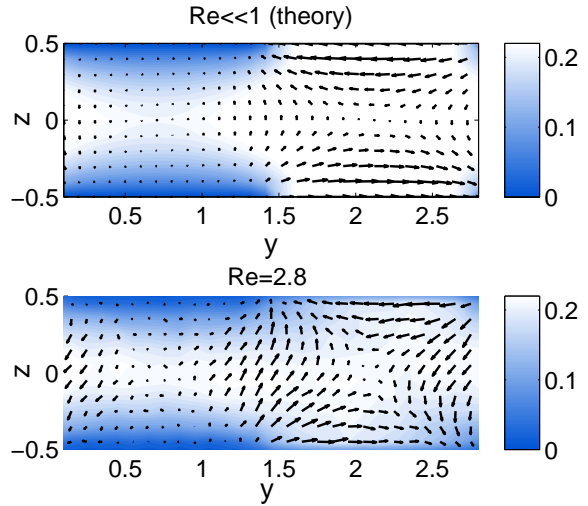}
   \caption{Cross-sections (at $x=0$) of the velocity field $(u_y, u_z)$ obtained in the theory (Re$\ll 1$) and simulations (Re$=2.8$). Color map shows the forward velocity $u_x$.
}
  \label{VelFields}
\end{figure}

Now we note that our simulations, performed at small, but finite Re, suggest that the texture misalignment leads to a complex shape of the flow on the scale of the
period of the texture. In Fig. \ref{VelFields} we have plotted typical theoretical (Re$\ll 1$) and simulation (Re=2.8) cross-sections of the fluid velocity. A general conclusion from this plot is that the fluid in the upper part of the channel is driven in the $y-$direction, but in the lower part it flows in the opposite direction, which is in agreement with our theoretical predictions of a transverse shear. A striking result emerging from this plot is that the simulated velocity field shows in addition a
significant $z-$component at the midplane of the channel, which is likely due to a finite inertia of the fluid.

\begin{figure}[t]
\includegraphics[width=1.\columnwidth]{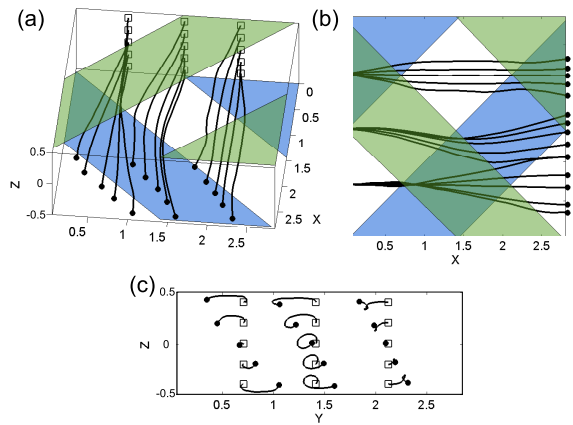}
   \caption{3D (a), top (b) and front (c) view of streamlines for the simulated velocity field (Re$=2.8$). Squares indicate initial positions of fluid particles, dots show their positions at the end of the periodic cell. }
\label{fig:strlines}
\end{figure}

To understand the global properties of the simulated flow, we now focus on its streamlines shown in Fig.~\ref{fig:strlines}. We conclude, that indeed, as expected, fluid particles are displaced in $y-$direction due to a generation of a non-zero mean shear. However, in addition some small (but non-zero)
vertical displacement is observed, so that they reach different positions on the $yz$-plane
after crossing a unit cell.

\begin{figure}[t]
\centering
\includegraphics[width=0.75\columnwidth]
{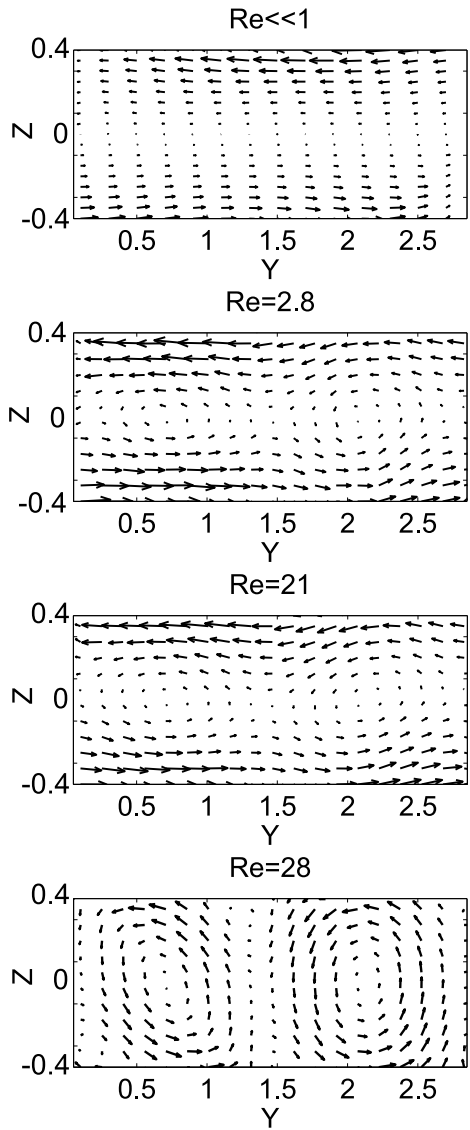}
   \caption{Vector field of fluid displacements over one cell period in $yz$-plane.
The velocity fields are obtained from simplified theory for $\mathrm{Re}\ll 1$
and simulations for $\mathrm{Re}=2.8$, $21$ and $28$.} \label{fig:dispfield}
\end{figure}

\begin{figure}[t]
\includegraphics[width=\columnwidth]
{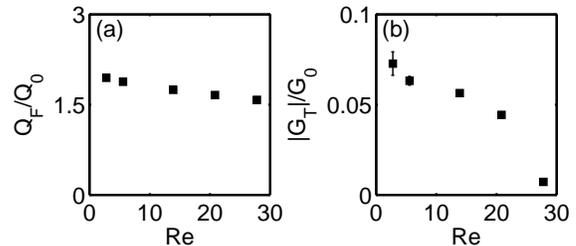}
   \caption{Forward flow rate (a) and transverse shear rate (b) simulated at different Re. } \label{fig:GTRe}
\end{figure}

An explanation for an emerging  flow in $z-$direction, not predicted by the simple Stokes theory, can be obtained if we invoke fluid inertia, which is characterized by the value of Re. To examine its significance in more details we construct a vector fields of fluid displacements over one period from the simulation data obtained at $\mathrm{Re}=2.8,\, 21$, and 28. In Fig.~\ref{fig:dispfield} we plot these  vector fields predicted theoretically and obtained from the simulations. We see that depending on the value of Re, different scenarios occur. For Re$\ll 1$ we observe a mean shear flow with very small vertical component near the walls. At $\mathrm{Re}=2.8$ and $21$ vector fields show a different 
behavior, i.e. represent a uniform shear superimposed with
two  co-rotating vortices. Finally, at $\mathrm{Re}=28$ we detect two isolated vortices and inhibited transport in $y$-direction.

The impact of Re on forward and transverse flows is highlighted in Fig.\ref{fig:GTRe}. We see that the simulated $Q_{\mathrm{F}}$ remains nearly constant.  In contrast, the transverse shear rate $|G_{\mathrm{T}}|$ is a relatively weakly decreasing function of Re up to $\mathrm{Re}\simeq 21$, and then decreases abruptly down to a much smaller value at $\mathrm{Re} = 28$.


\begin{figure}[h]
\includegraphics[width=\columnwidth]{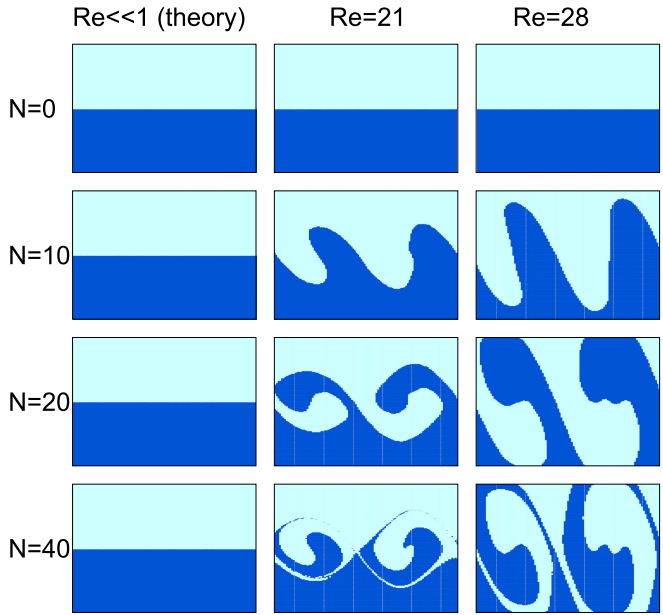}
   \caption{
Displacements of colored fluids over $N$ periods obtained for several Re. The axis limits (not shown) are the same as in Fig.~\ref{fig:dispfield}.  } \label{mixing}
\end{figure}

Our results have shown that flow in the channel with misaligned SH walls is much more complicated than the usual picture. The misalignment results in the generation of a transverse shear, but perhaps the most interesting and important aspect is the interplay of this transverse flow with the finite inertia of the fluid. This leads to a vertical displacement of a fluid, which could have implication for a vertical mixing in such a channel. To understand whether these results are significant for mixing, we mark the upper and lower parts of the channel by different colors and compute the evolution of this system \textbf{over} $N$ periods. The representative results for a contour field are shown in Fig.\ref{mixing}. We see that no vertical mixing occurs for the Stokes flow. For $\mathrm{Re}=21$ the mixing area is located in the center of the channel, but for $\mathrm{Re}=28$ the whole fluid is mixed.


\section{Conclusion}

\label{sec:conclusion}

In this paper we have extended the tensorial description of
SH channels developed earlier to configurations with
misaligned textures on bottom and top walls. We prove, that effective properties of
such channels are described by the permeability, $\mathbf k$, and the shear rate, $\boldsymbol\gamma$, tensors, and
develop a simplified theoretical approach, valid for Re $\ll 1$, to quantify them. We checked the validity of this approach by DPD simulations and found that theoretical predictions
are in quantitative agreement with the simulation results obtained at finite, but moderate Re.

Simulations have also shown that at finite Re the flow generated by this transverse
shear provides a basis for a new mixing mechanism of the fluids in microfluidic devices. Unlike
most classical mixers~\cite{rothstein.jp:2010,stroock2002b,mott2006}, where vortices are created both by large-scale inhomogeneities of patterned walls and by
the side walls of the channel, in our misaligned configuration the mixing occurs on the
scale of the texture period. It does not depend on the presence of confining side walls and is,
therefore, insensitive to the cross-sectional aspect ratio of the channel. We also recall that in microfluidic devices the mixing is normally controlled by viscous terms, so that the form of the flow is independent on Re and remains the same up to $\mathrm{Re}\simeq 100$ ~\cite{stroock2002a}. In contrast, our vertical mixing is driven by fluid inertia even at very moderate Re, and becomes more pronounced with the increase in Re.

Finally, we mention that our results can be immediately
extended to a situation of a channel with a periodically changing orientation of top and bottom stripes, which could lead to a chaotic advection~\cite{wiggins.s:2004,aref1984,stremler2004designing}, currently exploited mostly in  the `herringbone'-type mixers~\cite{stroock2002a,stroock2002b}.

\begin{acknowledgements}
This research was partly supported by the Russian Foundation for Basic Research (grant  15-01-03069), the EU
IRSES DCP-PhysBio (grant N269139),
and by the VW foundation.
The simulations were carried out using computational resources at the
John von Neumann Institute for Computing (NIC J\"ulich), the High Performance
Computing Center Stuttgart (HLRS) and Mainz University (MOGON).
\end{acknowledgements}

\appendix
\section{Calculation of permeability and shear tensors}
\label{appA}

We write the velocity field in the cell $\mathbf{u}(x,y,z)=\left(
u_{x},u_{y},u_{z}\right) $ in the form
\begin{equation}
\begin{array}{ll}
\displaystyle\mathbf{u}=\left\langle \mathbf{u}\right\rangle +\mathbf{u}_{1}+\mathbf{u}_{2}.    \label{u}
\end{array}
\end{equation}
Here $\mathbf{u}_{1},$ $\mathbf{u}_{2}$ are perturbations with zero mean
over the cell volume due to heterogeneous slippage at the
lower and upper walls, respectively.

Since the problem is linear, we can formulate the boundary conditions for
$\mathbf{u}_{1},$ $\mathbf{u}_{2}$ at the lower wall as
\begin{eqnarray}
z=-\frac{1}{2}:\quad \mathbf{u}_{2}=\mathbf{0},\quad u_{z1}=0,  \notag
\\
\mathbf{u}_{\tau 1}-b_{1}\frac{\partial \mathbf{u}_{\tau 1}}{\partial z}%
=-u_{sx}\mathbf{e}_{x}+\frac{\gamma _{yx}}{2}\mathbf{e}_{y}\\ \notag
+b_{1}\left(
\mathbf{e}_{x}+\gamma _{yx}\mathbf{e}_{y}+\frac{\partial \mathbf{u}_{\tau 2}%
}{\partial z}\right) ,
\label{bcu11}
\end{eqnarray}%
and at the upper wall as
\begin{eqnarray}
z=\frac{1}{2}:\quad \mathbf{u}_{1}=\mathbf{0,}\quad u_{z2}=0, \notag
\\
\mathbf{u}_{\tau 2}-b_{2}\frac{\partial \mathbf{u}_{\tau 2}}{\partial z}%
=-u_{sx}\mathbf{e}_{x}-\frac{\gamma _{yx}}{2}\mathbf{e}_{y}\\ \label{bcu2}
+b_{2}\left( -%
\mathbf{e}_{x}+\gamma _{yx}\mathbf{e}_{y}+\frac{\partial \mathbf{u}_{\tau 1}%
}{\partial z}\right) ,  \notag
\end{eqnarray}%
where $b_{1,2}\left( x,y\right) $ are the local slip lengths on the lower and
the upper wall, respectively.

The perturbations decay at a distance from the wall of the order of
$L$.  For this reason, in the case of a sufficiently thick channel we can expect that the
disturbance gradients at the opposite walls  are small:
\begin{eqnarray}
z &=&-\frac{1}{2}:\quad \left\vert \frac{\partial \mathbf{u}_{\tau 1}}{%
\partial z}\right\vert \ll \left\vert \frac{\partial \left\langle \mathbf{u}%
\right\rangle }{\partial z}\right\vert ,  \label{sm_dist} \\
z &=&\frac{1}{2}:\quad \left\vert \frac{\partial \mathbf{u}_{\tau 2}}{%
\partial z}\right\vert \ll \left\vert \frac{\partial \left\langle \mathbf{u}%
\right\rangle }{\partial z}\right\vert .  \nonumber
\end{eqnarray}

\begin{figure}
  \includegraphics[width=0.75\columnwidth]{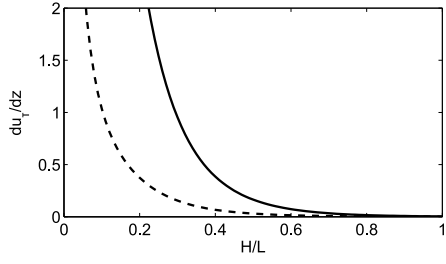}
  \caption{Maximal values of the velocity disturbance gradient at the opposite walls for $b/L=10$, $\phi=\delta/L=1/2$,  $2\alpha=\pi/2$  against the channel thickness. Solid curve shows $\dfrac{\partial u_{x1}}{\partial z}$, dashed curve plots $\dfrac{\partial u_{y1}}{\partial z}$ at $z=1/2$.}
  \label{fig:grads}
\end{figure}

Then the solution is the superposition
of two flows in a channel with a SH and a no-slip walls. These  flows can be
calculated numerically~\cite{nizkaya.tv:2013}. The validity of the assumption, Eq.(\ref{sm_dist}),
is verified by evaluating the disturbance gradients induced by a SH wall on a
no-slip wall. The gradients are small compared to the wall gradient for the
Poiseuille flow, $\left\vert \partial \left\langle \mathbf{u}\right\rangle
/\partial z\right\vert =2$, at $H/L>0.5$ (see Fig.~\ref{fig:grads}).

\begin{figure}
  \includegraphics[width=0.5\columnwidth]{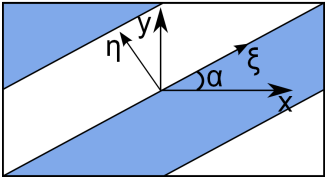}
  \caption{Coordinate system $(\xi,\eta)$, associated with the lower wall, see Eq.(\ref{bas}).}
  \label{fig:sketch3}
\end{figure}

Consider now the flow induced by the undisturbed
shear rate $\mathbf{e}_{x}+\gamma _{yx}\mathbf{e}_{y}$ at the bottom wall.
Projections of the unit vectors to the $\left( \xi ,\eta \right) $
coordinate system with the axes parallel and perpendicular to the stripes (see Fig.~\ref{fig:sketch3}) are%
\begin{equation}
\begin{array}{lll}
\mathbf{e}_{x} &=&\mathbf{e}_{\xi }\cos \alpha -\mathbf{e}_{\eta }\sin
\alpha ,  \label{bas} \\
\mathbf{e}_{y} &=&\mathbf{e}_{\xi }\sin \alpha +\mathbf{e}_{\eta }\cos
\alpha .
\end{array}
\end{equation}

It follows from Eqs. (\ref{bcu11}) and (\ref{bas}) that at $z=-1/2$ the mean slip velocity and the shear
rate  are%
\begin{eqnarray}
\left\langle \mathbf{u}_{\tau }\right\rangle  &=&\left( u_{sx}\cos \alpha -%
\frac{\gamma _{yx}}{2}\sin \alpha \right) \mathbf{e}_{\xi }  \notag \\
&&-\left( u_{sx}\sin \alpha +\frac{\gamma _{yx}}{2}\cos \alpha \right)
\mathbf{e}_{\eta },  \label{avu} \\
\left\langle \frac{\partial \mathbf{u}_{\tau }}{\partial z}\right\rangle
&=&\left( \cos \alpha +\gamma _{yx}\sin \alpha \right) \mathbf{e}_{\xi
}-\left( \sin \alpha -\gamma _{yx}\cos \alpha \right) \mathbf{e}_{\eta }.
\notag
\end{eqnarray}%
From (\ref{avu}) we have%
\begin{equation}
\frac{b_{\mathrm{eff}}^{\Vert }}{H}=\frac{u_{sx}\cos \alpha -
\frac{\gamma_{yx}}{2}\sin \alpha }{\cos \alpha +\gamma _{yx}\sin \alpha },
\label{b_eff}
\end{equation}
\begin{equation}
\frac{b_{\mathrm{eff}}^{\perp }}{H}=\frac{u_{sx}\sin \alpha +\frac{\gamma
_{yx}}{2}\cos \alpha }{\sin \alpha -\gamma _{yx}\cos \alpha }.
\label{b_eff2}
\end{equation}
Here the components of the effective slip length tensor $\mathbf{b}_{\mathrm{eff}}$
correspond to the usual configuration with a SH and a no-slip walls~\cite{harting.j:2012}. Thus
we obtain from (\ref{b_eff}), (\ref{b_eff2}):
\begin{eqnarray}
u_{sx}(\alpha)=\frac{\beta_++\beta_-\cos(2\alpha)+2(\beta_+^2-\beta_-^2)}{1+2\beta_+-2\beta_-\cos(2\alpha)},\\
\gamma_{yx}(\alpha)=-\frac{2\beta_-\sin(2\alpha)}{1+2\beta_+-2\beta_-\cos(2\alpha)},
\end{eqnarray}%
where $\beta_{+}=(b_{\mathrm{eff}}^{\Vert }+b_{\mathrm{eff}}^{\perp })/(2H)$, $\beta_{-}=(b_{\mathrm{eff}}^{\Vert }-b_{\mathrm{eff}}^{\perp })/(2H).$ 


The flow in the transverse direction can be obtained by replacing $x$ by $y$ and $\alpha$ by $\pi/2-\alpha$.
Therefore, we readily obtain the permeability tensor:
\begin{equation}
\mathbf{k}=\left(
\begin{array}{cc}
1+6u_{sx}(\alpha) & 0 \\
0 & 1+6u_{sx}(\pi/2-\alpha)%
\end{array}%
\right) ,
\label{ks}
\end{equation}
and the shear tensor:
\begin{equation}
\boldsymbol\gamma=\left(
\begin{array}{cc}
0 & \gamma _{yx}(\pi/2-\alpha) \\
\gamma _{yx}(\alpha) & 0%
\end{array}%
\right) .
\label{gam}
\end{equation}

\bibliography{crossed}

\end{document}